\input ./psfig.tex
\documentstyle[]{mn14}
\def\mch{M$\rm^{c}$Hardy~}
\def\etal{\it et al.~\rm}

\title[X-ray QSO evolution]
{X-ray QSO evolution from a very deep ROSAT survey}

\author[Jones, L.R. \etal]
{
L R Jones$^{1,2}$, I M M$\rm^{c}$Hardy$^{2}$, M R Merrifield$^{2}$, 
K O Mason$^{3}$, P J Smith$^{3}$, \and R G Abraham$^{4}$, 
G Branduardi-Raymont$^{3}$, 
A M Newsam$^{1}$, G Dalton$^{5}$, \and M Rowan-Robinson$^{6}$,
G Luppino$^{7}$ \\
$^{1}$ Code 660.2, NASA/Goddard Space Flight Center, Greenbelt, MD 20771, USA.\\
$^{2}$ Department of Physics, The University, Southampton SO17 1BJ, UK.\\
$^{3}$ Mullard Space Science Laboratory, University College London, 
Holmbury St Mary, Dorking RH5 6NT, UK.\\
$^{4}$ Institute of Astronomy, University of Cambridge, Madingley Rd, Cambridge CB3 
OEZ, UK.\\
$^{5}$ Department of Astrophysics, University of Oxford, Keble Road, Oxford OX1 3RH, UK.\\
$^{6}$ Astrophysics Group, Blackett Laboratory, Imperial College, London SW7 2BZ, UK.\\
$^{7}$ Institute for Astronomy, University of Hawaii, 2680 Woodlawn Drive, Honolulu, 
 Hawaii 96822, USA.\\
}
\begin{document}
\maketitle

\begin{abstract}
In the deepest optically identified  X-ray survey yet performed, we
have identified 32 X-ray selected QSOs to a flux limit of $2 \times 
10^{-15}$ erg cm$^{-2}$ s$^{-1}$ (0.5-2 keV).  The survey, performed with the 
ROSAT PSPC, has 89\% spectroscopic completeness. The QSO log(N)-log(S) 
relation is found to have a break to a flat slope at faint fluxes.
The surface density of QSOs at the survey limit is 230$\pm$40  per 
square degree, the largest so far of any QSO survey.
We have used this survey to measure the QSO X-ray 
luminosity function at low luminosities (L$_{X}<10^{44.5}$ erg s$^{-1}$)
and high redshifts (1$<$z$<$2.5). The highest redshift 
QSO in the survey has z=3.4.
Combined with the QSOs from the {\it Einstein} EMSS at bright fluxes,
we find pure luminosity evolution of the form L$_{X}\propto$(1+z)$^{3.0(+0.2,-0.3)}$ is an 
adequate description of 
the evolution of the X-ray luminosity function at low redshifts. A redshift cutoff in the
evolution is required at z=1.4 $^{+0.4}_{-0.17}$ (for q$_{0}$=0.5). We discuss the
form of this evolution, its dependence on the model assumed and the errors on the
derived parameters. We show that most previous X-ray surveys, including the EMSS,
are consistent with a power law luminosity evolution index of 3.0. 

The contribution of QSOs to the 1-2 keV cosmic X-ray background is 
found to be between 31\% and 51\%.  

\end{abstract}


\section{INTRODUCTION}
The evolution of the Active Galactic Nuclei population can be studied
up to redshifts corresponding to $\approx$90\% of the age of the Universe,
giving information on the nature of the AGN themselves and on the nature
of the Universe at early epochs. Optical surveys (e.g. Boyle \etal 1991)
have shown that pure luminosity evolution, in which the luminosity function
simply moves to higher luminosities at higher redshifts, is a good 
description of QSO evolution at redshifts 0.3$<$z$<$2. However, in the 
surveys
of Hewett \etal (1993) and Miller \etal (1993) there is in addition a 
change in the slope 
of the luminosity function with redshift, such that the most luminous QSOs
show no evidence for any evolution.

Optically, 
QSOs are proportionally very rare objects which have to be carefully selected
from the overwhelming number of galaxies and stars,
usually using colour and morphological criteria based on well calibrated 
photographic plates. 
This selection of stellar
objects produces incompleteness at low redshift, because the host galaxy 
may be resolved,
and at low AGN luminosities, when the host galaxy  may also contaminate 
the AGN colours.
In contrast, X-ray surveys of AGN are very direct, since
AGN are the most numerous type of source in
current X-ray surveys.

The {\it Einstein} EMSS survey contains 427 QSOs with a median redshift of 0.3
(Maccacaro \etal 1991; Della-Ceca \etal 1992). (We use the term QSOs to 
refer to all
broad line objects, including Seyfert 1 AGN and QSOs). ROSAT pointed
surveys reach to fainter fluxes and higher redshifts.
The RIXOS survey of
Page \etal (1996) has a limiting flux of 3x10$^{-14}$ erg cm$^{-2}$ 
s$^{-1}$ (0.5-2 keV) and
the survey of Boyle \etal (1994) reaches 4x10$^{-15}$ erg cm$^{-2}$ 
s$^{-1}$.
The results presented here are based on
the UK ROSAT deep field survey. This is currently the faintest optically 
identified X-ray survey,
 reaching a flux limit of 2x10$^{-15}$ erg cm$^{-2}$ s$^{-1}$, and is thus 
sensitive to QSOs of
low luminosity (L$_{X}{\sim}10^{44}$ erg s$^{-1}$) at redshifts up to z=2.
The general properties of all the X-ray source populations
detected in this survey, including the numerous narrow emission line 
galaxies detected at faint fluxes, are described
in  \mch\etal (1996). The log(N)-log(S) relation of all X-ray sources
combined has been investigated by Branduardi-Raymont \etal (1994)
and Barcons \etal (1995). Romero Colmenero \etal (1996) have investigated the
ROSAT X-ray spectra of the sources in this survey and shown that 
the mean X-ray spectral index of the QSOs is 0.92$\pm$0.02.

In section 2 we describe the X-ray and optical observations. In the 
following sections we explain our analysis technique and describe the 
results. In section 5 we discuss  these results. Section 6 lists
the conclusions.

\section{OBSERVATIONS}

\subsection{X-ray Observations~}
Full details of both the X-ray and optical observations are given in \mch\etal
1996. Here we summarize the main points. 
The survey field was selected for its low Galactic absorption, and
ROSAT Position Sensitive Proportional Counter (PSPC) observations made at two epochs,
giving a total exposure of 111 ksec.
The inner 15 arcmin radius area of the combined
exposure  was searched for
point-like X-ray sources. We used our own source detection 
software, employing a maximum-likelihood fit to the point spread 
function (Cash 1979) at each pixel of size 5 arcsec, and set 
the detection threshold so that one false source would be 
detected in the survey area. 
The increase in 
size of the point spread function (PSF) with 
off-axis angle, from a full width at half maximum (FWHM) of 25 arcsec on axis 
to a FWHM of 58 arcsec at 15 arcmin off axis (at an energy of 1 keV; Hasinger \etal 1993b),
limited the survey to a radius of 15 arcmin. The 
0.5-2 keV band was used to detect sources,
 rather than the full 0.1-2.4 keV ROSAT PSPC band, in order to 
reduce the background level 
and minimize the size of the instrumental PSF.
There are 105 X-ray sources above 1.6x10$^{-15}$ erg cm$^{-2}$ s$^{-1}$
in the 30 arcmin diameter survey region.
We set the flux limit for a complete sample higher, at 
2x10$^{-15}$ erg cm$^{-2}$ s$^{-1}$ (0.5-2 keV),
corresponding to 3.8$\sigma$ confidence. 
Simulations showed that at this flux limit
a log(N)-log(S) relation of the form measured by Branduardi-Raymont \etal
(1994) could be successfully recovered, and thus that source confusion
was not a major problem.

The limiting X-ray flux is a function of off-axis angle because of the 
strong increase in size of the PSF even within our survey region of
15 arcmin radius  and because of mirror vignetting. 
The limiting flux was calculated
in annuli of size 0.5 arcmin, including the PSF at 1 keV given by Hasinger \etal (1993b),
the mirror vignetting, the background level and the shape of our survey area (two
outer regions of the 15 arcmin radius circular area were not covered in our
spectroscopic follow-up).
The area of the annuli were then summed to give 
the total area surveyed as a function of flux. 
At fluxes $>$2.7x10$^{-15}$ erg cm$^{-2}$ s$^{-1}$ (0.5-2 keV) the area surveyed was
0.16 sq deg. The area surveyed at fainter fluxes is given in the second column of 
Table 1.

\begin{table}
\centering
\caption {Area surveyed as a function of flux.}

\begin{tabular}{lll} \hline
Flux x10$^{-15}$ & Sky area & Effective sky area \\
erg cm$^{-2}$ s$^{-1}$ &  & after correction for \\
(0.5-2 keV) & deg$^{2}$ & incompleteness \\
            \hline
 & & \\
2.0 & 0.062 & 0.055 \\
2.1 & 0.084 & 0.075 \\
2.2 & 0.104 & 0.094 \\
2.3 & 0.119 & 0.106 \\
2.4 & 0.133 & 0.118 \\
2.5 & 0.145 & 0.129 \\
2.6 & 0.157 & 0.140 \\
2.7 & 0.159 & 0.140 \\
3.6 & 0.159 & 0.147 \\
4.2 & 0.159 & 0.154 \\
4.3 & 0.159 & 0.159 \\
\end{tabular}
\end{table}

\subsection{Optical Observations~}
Optical CCD imaging in V,R \& I bands has been obtained at
the University of
Hawaii 88inch telescope, 2.5m Nordic Optical Telescope, 3.6m Canada-France-Hawaii
Telescope (CFHT) and Michigan-Dartmouth-MIT  2.4m telescope. 
The deepest images reach at least R=24.5 mag. We have also made deep VLA radio 
maps at 20cm and
6cm, reaching a flux limit of 0.5 mJy at 20 cm. 
The ROSAT positions were 
corrected for the small ROSAT PSPC systematic position error
using three independent methods (bright star/ROSAT coincidences, VLA/ROSAT coincidences, and
the first few bright AGN spectroscopic identifications)
which all gave a consistent
offset of size 13 arcsec. A correction for the small ROSAT roll angle error of 0.185 
degrees (Briel \etal 1995) was also applied.
The remaining random error was 10 arcsec at 95\% confidence.
Most (90\%) of the X-ray sources were identified with objects brighter than R=23 mag
and $\approx$30\% had two or more possible counterparts of R$\leq$22.5 mag.

Low resolution spectroscopy (10-15\AA) was performed at the 3.6m CFHT with the
MOS multislit spectrograph and at the 4.2m William Herschel Telescope (see 
\mch\etal
1996 for details). Spectra were obtained within a contiguous region 
containing 73 X-ray sources above the  flux limit of 
2x10$^{-15}$ erg cm$^{-2}$ s$^{-1}$ (0.5-2 keV). The survey region was largely defined
by the positioning of the MOS fields and did not include some areas at large off-axis angles.
Spectra were obtained of most objects of stellar appearance and R$<$22.5 mag in 
the error circles,
irrespective of other possible counterparts, in an attempt
to identify all the X-ray sources which could be QSOs. The spectra of R=22.5 mag objects were
of sufficient signal/noise ($\approx$10) to identify typical QSO broad emission lines.
In general, optical colours were not used to select candidates for spectroscopic observations
in cases where two or more candidates existed, in order not to bias the identifications.
In 9 of the 32 error circles in which QSOs were found a second object of
R$\la$22.5 mag
was also present. In 3 cases the second object had a stellar appearance. Spectra were
obtained of 2 of the 3 stellar objects, showing them to be normal Galactic stars. In 
all three cases the QSO was taken as the 
counterpart since (a) the QSO was nearer the
error circle centre and (b) if all the X-ray flux was assumed to come from the star,
its X-ray/optical flux ratio fell beyond the range found for stars of the same stellar type in
the EMSS by Stocke \etal (1991). 
In six cases the second object was a faint galaxy, and since 
in five of them the QSO was nearest the error circle 
centre, and the X-ray/optical flux ratios
of the QSO fell within the range found in the EMSS for AGN, 
it was taken as the counterpart. In one case
(object 37) a faint galaxy of R=22.5 mag is 2 arcsec from the error circle centre and the QSO is
7.3 arcsec away. Since the 
total number of unrelated galaxies in all 32 QSO error circles, predicted from 
the optical counts of Metcalfe \etal (1991), is 8 at R$<$22 mag 
and 13  at R$<$22.5 mag, we have assumed 
that all these  six galaxies are unrelated to the X-ray emission.

In 2 of the 32 X-ray sources another point X-ray source is closer than 1 arcmin. In these
cases the X-ray source centroid positions produced by the source detection 
algorithm were slightly distorted and the correct positions
were determined by an inspection of the X-ray image and overlaying the X-ray and optical images.
In one case 
both sources (sources 29 and 154) are identified with QSOs 50 arcsec apart, but at 
different redshifts. In the
other case (source 23) the QSO, at a redshift of z=0.97, and a narrow emission line galaxy
at a redshift of z=0.18, are separated by only 20 arcsec. The peak of the X-ray
emission lies nearest the QSO, and we have ignored the contribution from the narrow 
emission line galaxy. The X-ray/optical position offsets have been investigated omitting these
three sources (numbers 23, 29 and 154) and splitting the sample into two flux ranges, with
equal numbers of QSOs in each range.
The mean position error for bright QSOs with
flux $>$8x10$^{-15}$ erg cm$^{-2}$ s$^{-1}$ (0.5-2 keV) was 3.0 arcsec (and within 5 arcsec for 95\%
of them), and for 
faint QSOs with flux $<$ 8x10$^{-15}$ erg cm$^{-2}$ s$^{-1}$ (0.5-2 keV) it was 5.2 arcsec 
(and within 10 arcsec for 95\% of them),
confirming the error circle size.

The 32 X-ray sources (out of a total of 73) with counterparts having at least one 
emission line of FWHM$>$1000 km s$^{-1}$ have been
included in this
analysis. Reliable redshifts from two or more emission lines were obtained
for 90\% of the QSOs. Eight sources out of 73 (11\% of the total) were unidentified. The 
error circles of
three of these contain galaxies of R$\approx$21-22 mag with spectra of insufficient  
quality to classify them. The remaining five error circles are blank to R=23 mag, and
deeper imaging has revealed only faint galaxies. We have assumed that
the unidentified sources contain the same fraction of QSOs as the identified sources
at the same flux 
and reduced the sky area as a function of flux by the fraction of all sources above
each flux that have been identified. 
The corrected sky area is listed in Table 1.
Since the unidentified
sources are not among the brightest sources in the survey, 
and the QSO fraction of the identified sources falls with
flux, the correction for incompleteness is not large. It 
is the equivalent of assuming 3 of the 10 unidentified sources are QSOs,
in addition to the 32 identified QSOs.
In practice, this correction may be too large, since QSOs, with their
strong, broad emission lines, are the easiest
class of X-ray source to identify.

\section{Analysis}
\subsection{X-ray spectra and the EMSS}
Galactic absorption in the direction of this field is uniform and
low (N$_H$=6.5x10$^{19}$ cm$^{-2}$, Branduardi-Raymont  \etal 1994). The small variations in
the Galactic column density ($\sim$2x10$^{19}$ cm$^{-2}$) will have negligible effect 
on the flux at energies $>$0.5 keV, as used here. Thus 
ROSAT PSPC count rates were converted to 0.5-2 keV fluxes assuming a
fixed column density of 6.5x10$^{19}$ cm$^{-2}$ and a 
power law spectrum of energy index 1. The conversion factor was 1.1x10$^{-11}$ 
erg cm$^{-2}$ s$^{-1}$ (ct s$^{-1}$)$^{-1}$. Romero Colmenero \etal (1996)
have shown that a power law  is a good
description of most ($\approx$75\%) of the spectra of the QSOs in this survey.
The mean spectral index in the 0.1-2 keV band, ignoring the brightest ten QSOs,
is 0.96$\pm$0.03. If all QSOs are summed into an average spectrum,
the mean spectral index is 1.21$\pm$0.03 (and is dominated by the brightest few 
QSOs) and the conversion factor used above would
change by $<$1\%.

The ROSAT survey was analysed in combination with the EMSS survey of
Stocke \etal (1991). The EMSS survey
defined the low redshift X-ray luminosity function (XLF) and the high luminosity part of 
the high redshift XLF.
In order to combine the two surveys, 
the ROSAT fluxes were converted to the {\it Einstein} 0.3-3.5 keV band via a constant
conversion factor of 1.8. This factor is accurate to within 10\% for simple power
law spectra of energy index 0.3-2, assuming zero absorption intrinsic to each source
(see Section 3.4 for the results of using a different conversion factor).
All but one of the QSO spectral indices observed in the ROSAT sample are consistent with 
lying within this range (Romero Colmenero \etal 1996, figure 1).
The 427 EMSS  QSOs (Stocke \etal 1991) were used together with the 21 `expected' QSOs of
Maccacaro \etal (1991), to account for EMSS incompleteness. The EMSS AGN sample 
included 32 sources which have an ambiguous classification and/or are uncertain 
identifications (tables 8 and 10 of Stocke \etal 1991). The EMSS optical 
spectra of these objects
generally showed evidence of high ionization levels ([OIII]$\geq$[OII]) but either 
the signal/noise was 
insufficient to detect a broad permitted line component, or there was no
coverage of H$\alpha$. Spectra of higher resolution and higher signal/noise of some 
of these objects have been obtained by Fruscione \& Griffiths (1991),
Boyle \etal (1995) and Halpern, Helfand \& Moran (1995). Whilst in some of the objects
the X-ray emission may arise from star formation activity, many of them
do have weak broad permitted lines and may thus harbour AGN. Until detailed spectroscopy
is available for all these objects, we have included them all in the analysis.
They are all at redshifts z$\leq$0.42.

Similar objects have been detected in our ROSAT survey. Initial spectroscopy
of the sixteen narrow emission line galaxies indicates that they are also probably a
mixture of starburst and Seyfert galaxies, at redshifts z$\leq$0.59 (\mch\etal 1996).
We determine the effect of including these objects in the ROSAT sample on the 
XLF analysis below.

\subsection{Redshift, Luminosity and Flux Distributions}

The redshift distribution of the QSOs in the ROSAT survey is shown as a
solid line
in Figure 1.  The EMSS QSO redshift distribution is shown as a
dotted line. The median ROSAT redshift is z=1.6, much higher than the median    
redshift in the EMSS of 0.3.
There is a decline in the number of ROSAT QSOs starting at a redshift of
z$\approx$1.8.
In Figure 2 the same ROSAT QSO redshift distribution is plotted, together with
the redshift distribution of the sum of the ROSAT QSOs and narrow emission line
galaxies (NELGs) as a dashed line.
The difference in the redshift distributions of the two types of source
immediately
suggests that two different source populations are involved, even though the    
X-ray emission
in some fraction of the NELGs may arise in Seyfert II AGN. We return to this 
point in the discussion.

\begin{figure}
\psfig{figure=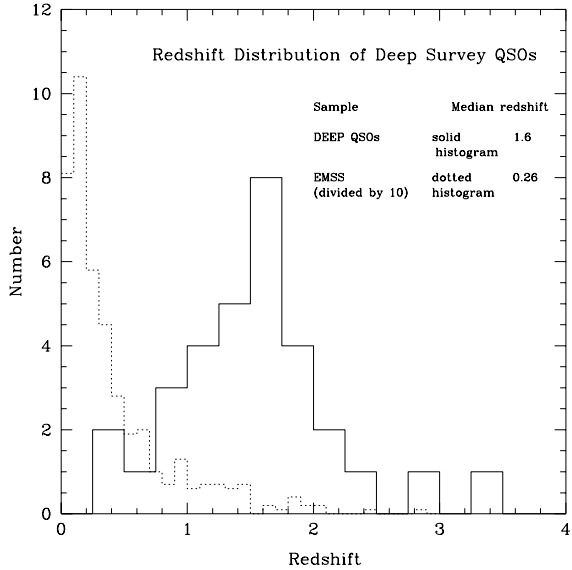,height=8.0truecm,width=8.0truecm,angle=0}
\caption{ Redshift distribution n(z) of the QSOs in the ROSAT Survey (solid line),
and QSOs in the EMSS (dotted line). 
The ordinate is the number of objects per redshift interval of size 0.25. 
The EMSS numbers have been divided by a factor of 10 to fit on the plot.}
\end{figure}

\begin{figure}
\psfig{figure=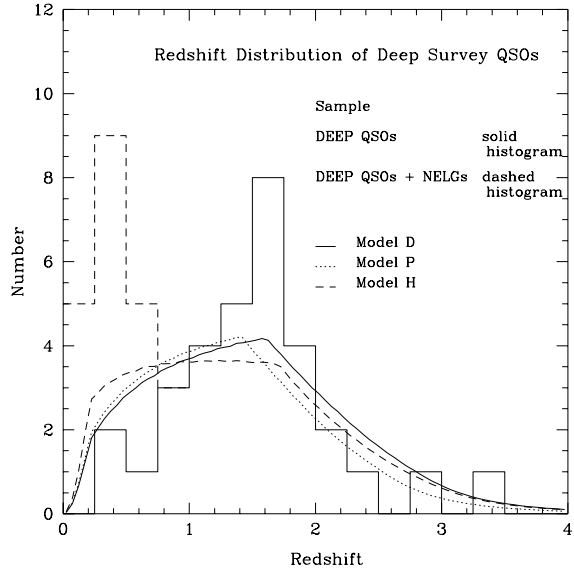,height=8.0truecm,width=8.0truecm,angle=0}
\caption{ 
Redshift distribution of the QSOs in the ROSAT survey as in Figure 1, together
with the sum of the ROSAT 
QSOs and narrow emission line galaxies (dashed line), and the predicted n(z)
of models D, P \& H. (see text). Note that the dashed histogram follows the solid
histogram exactly at z$>$0.75, because there are no nelgs at z$>$0.75.
}
\end{figure}

The X-ray luminosity-redshift diagram is shown in
Figure 3, for the ROSAT sample (solid symbols) and for the EMSS sample (open symbols).
Also shown are lines of constant flux corresponding to the approximate limiting fluxes of
the two surveys. The rest frame 0.3-3.5 keV luminosities were calculated assuming 
H$_{0}$=50 km s$^{-1}$ 
Mpc$^{-1}$, q$_{0}$=0.5 and an X-ray spectral index of 1.  This spectral index
is the mean value for radio quiet QSOs in the EMSS (Wilkes \& Elvis 1987).
The ROSAT survey samples
luminosities a factor $\approx$30 lower than the EMSS survey at redshifts z$>$1.
A QSO of luminosity $\sim$10$^{44}$ erg s$^{-1}$, typical for nearby QSOs
(Maccacaro \etal 1991), is detectable up to a redshift of z=2 in the ROSAT
survey, and
we thus expect the ROSAT survey to help define the low luminosity, high redshift 
part of the luminosity function. At redshifts z$>$1.5 there are approximately
equal numbers of QSOs in the ROSAT survey as in the EMSS, but at very different
luminosities. 

\begin{figure}
\psfig{figure=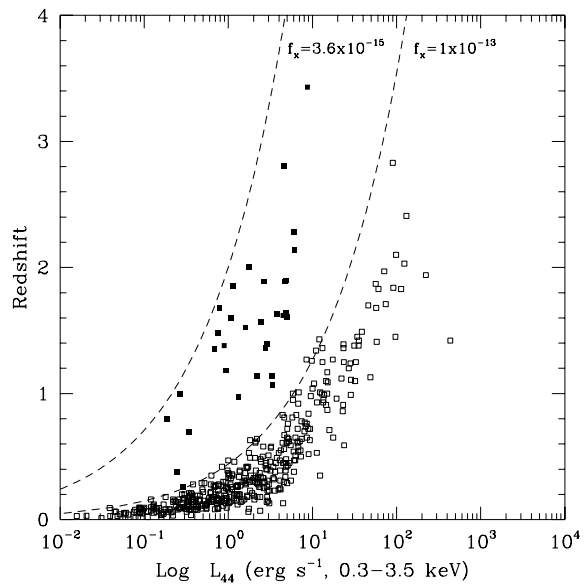,height=8.0truecm,width=8.0truecm,angle=0}
\caption{ 
X-ray luminosity-redshift diagram for ROSAT QSOs (solid symbols) and EMSS 
QSOs (open symbols). Lines of constant flux corresponding to the approximate survey
flux limits are also shown.
}
\end{figure}

The QSO integral number-flux relation, or log(N)-log(S) diagram, is shown in Figure 4a.
The ROSAT QSO fluxes have been converted to the {\it Einstein} 0.3-3.5 keV band
using the conversion factor of 1.8 described above. A correction has
also been made for the QSOs at bright fluxes missed in the ROSAT pencil-beam survey
because of their rarity on the sky. This small correction was 6 deg$^{-2}$,
corresponding to one QSO predicted in the survey area (at fluxes of 
$>$9x10$^{-14}$ erg cm$^{-2}$ s$^{-1}$, 0.3-3.5 keV). 
The surface density of QSOs at the flux limit of the ROSAT survey is 230$\pm40$
deg$^{-2}$, the largest measured for a QSO survey at any wavelength.
The counts are in good agreement with those of Boyle \etal (1994) (shown as 
filled circles in Figure 4a), except for a slight excess in our field at
a flux $\approx$2x10$^{-14}$ erg cm$^{-2}$ s$^{-1}$.

A break in the integral QSO counts of Figure 4a is apparent at a 
flux $\approx$2x10$^{-14}$ erg cm$^{-2}$ s$^{-1}$
(0.3-3.5 keV). This break can also be clearly seen in the differential log(N)-log(S) relation
of Figure 4b. The slope of the QSO log(N)-log(S) relation, measured from the 21
QSOs fainter than this flux, is -0.3$\pm$0.4. The slope was measured from a 
minimum $\chi^{2}$ fit to the differential counts. It is significantly flatter
than the slope of -1.61$\pm$0.06 measured at bright fluxes in the EMSS QSO sample
by Della-Ceca \etal (1992), at $>3\sigma$ significance. 
The break in the QSO log(N)-log(S) relation occurs at approximately the same flux
as the break in the counts of all X-ray sources 
measured by Branduardi-Raymont \etal (1994). Since QSOs account for $\approx$80\%
of all X-ray sources at the break flux (Shanks \etal 1991, \mch\etal 1996),
it is the QSOs which are responsible for this break.
Although there is no overlap
in the flux range of the QSOs in the ROSAT survey and the EMSS, the bright ROSAT QSO
counts are consistent with a direct extrapolation of the
EMSS QSO counts, given by the dotted line in Figure 4a.
However, since there are only eleven ROSAT QSOs at fluxes brighter than 
the break, the count slope and normalisation at bright fluxes are not well constrained.


\begin{figure}
\psfig{figure=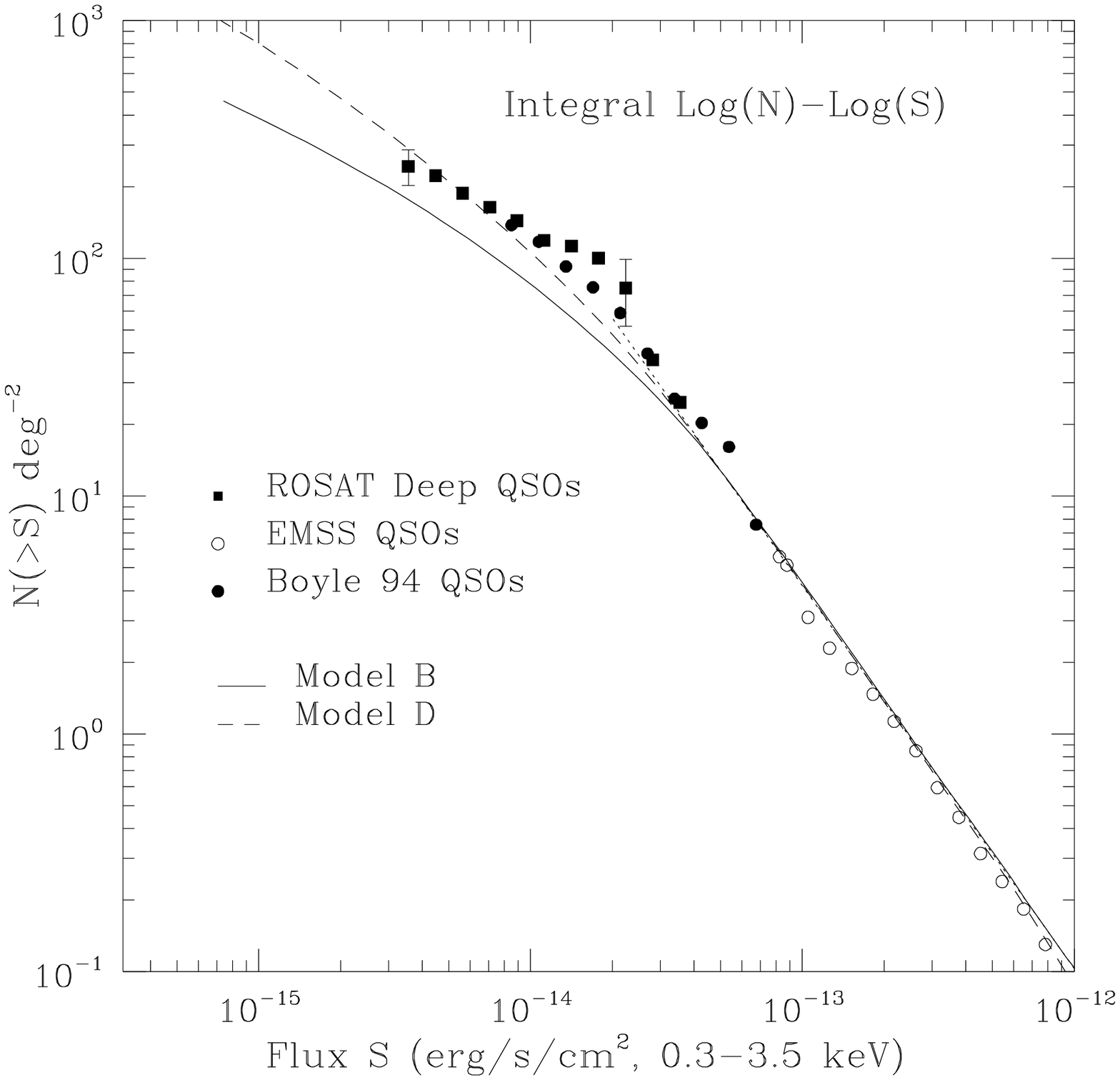,height=8.0truecm,width=8.0truecm,angle=0}
\psfig{figure=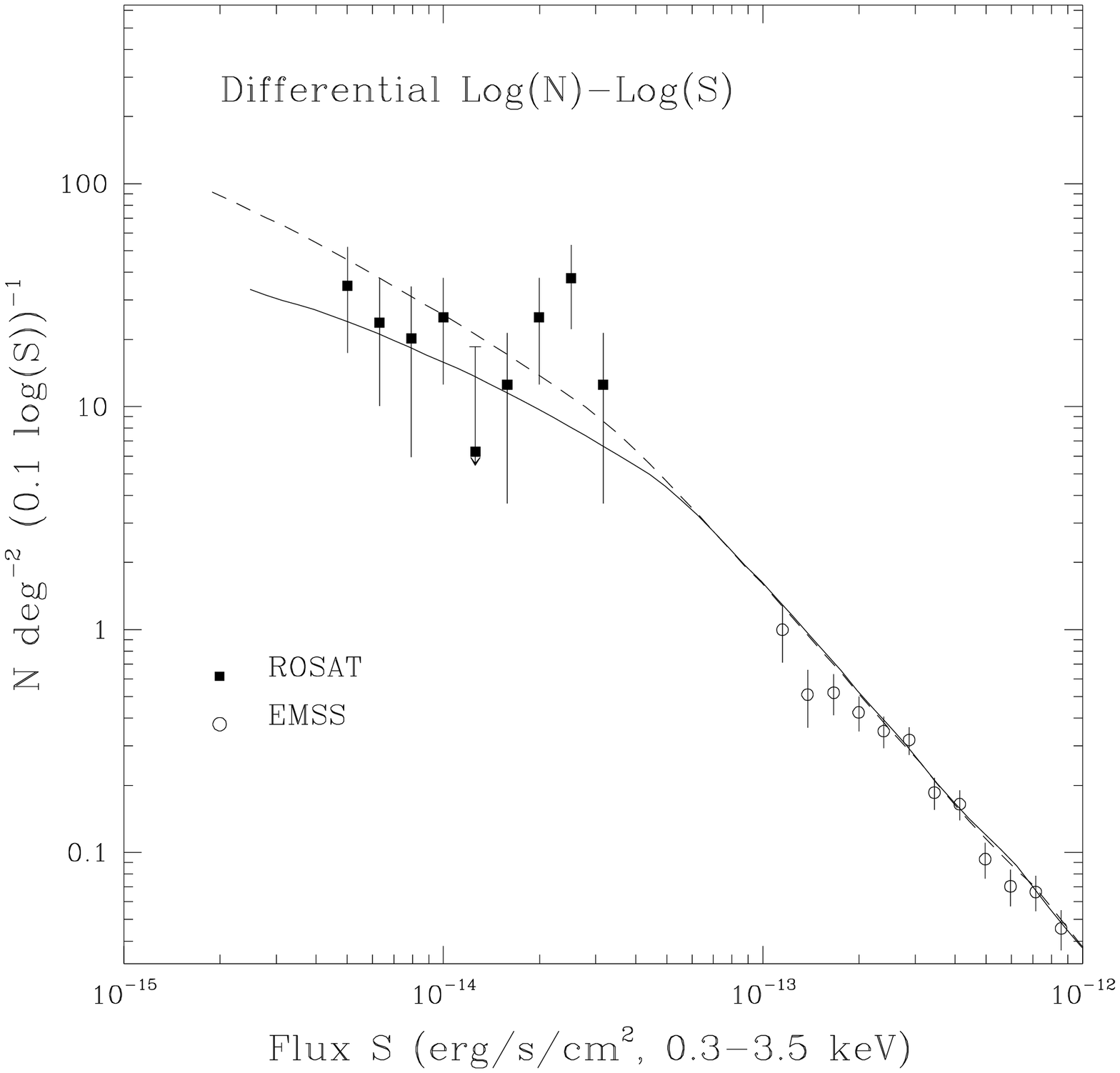,height=8.0truecm,width=8.0truecm,angle=0}
\caption{ 
(a) Integral log(N$>$S)-log(S) relation for QSOs in the ROSAT deep survey (filled 
squares), EMSS AGN (open circles), and Boyle \etal (1994) ROSAT survey (filled 
circles, shifted by 0.03 in log(S) for clarity).
The dotted line shows the extrapolation of the
best fit EMSS slope of 1.61 to fainter fluxes. Models B (solid line) and D (dashed line)
are described in the
text. (b) the same data as in (a) are plotted, but in independent, differential bins.
}
\end{figure}

\subsection{The X-ray Luminosity Function}
\subsubsection{The binned Luminosity Function}
An initial estimate of the differential  X-ray luminosity function (XLF) was obtained
by binning the QSOs from the combination of the two
surveys in redshift and luminosity. We used the 1/$V_{a}$  statistic 
of Avni and Bachall (1980) as described by Maccacaro \etal (1991).
The XLF at
five redshift intervals is shown in Figure 5a for $q_{0}=0.5$ and Figure 5b
for $q_{0}=0$. 
The XLF is plotted here per 
logarithmic luminosity bin (although it is defined below per linear luminosity bin) 
so that pure luminosity evolution will produce a simple shift along the luminosity 
axis. 
K-corrections were unity, assuming a spectral index of 1.
Error bars have been estimated assuming Poissonian errors based on the 
number of QSOs in each bin. All points represent at least two QSOs. Upper
limits to the XLF have been plotted for some bins containing one or zero QSOs.
The upper limit was set at 3 QSOs, corresponding to 80\% confidence.
Where there was one QSO detected,
the position of the symbol represents the value of $\Phi$ given by
the single QSO, and the upper limit represents $<$3 QSOs.
The lowest three redshift intervals are those used by Boyle \etal (1994) and
give a constant width in log(1+z). The remaining redshift intervals 
explore the XLF behaviour at the redshift z$\approx$2 where 
a halt or slowing of the pure luminosity evolution is found in the optical surveys
of Boyle \etal (1991), although because these binned XLF estimates do not take into
account any evolution occurring within each bin, they are not used to 
quantify the 
detailed XLF evolution.

The XLF has a two power law form at all redshifts where there are 
enough QSOs to provide a good measurement (up to z=2.2). 
Comparison with Figure 6 of Maccacaro \etal (1991) shows that the addition 
of the ROSAT data has extended the measurement of the XLF to low
luminosities 
(lower than the break luminosity) at redshifts  0.4$<$z$<$2.2. Our
measurement of the XLF also extends to lower luminosities than
the ROSAT surveys of Boyle \etal (1994) and Page \etal (1996). As noted in
previous investigations, the general form of the evolution of
the XLF is of luminosity evolution, in which the shape of the
XLF and the  number of QSOs 
are conserved with redshift, but the characteristic QSO luminosity 
increases with redshift. For q$_0$=0.5, it is clear from Figure 5a that 
the luminosity evolution which appears to
be a good description at low redshifts does not continue to high redshifts.
The XLFs  at redshifts 1.8$<$z$<$2.2 and 2.2$<$z$<$3.5
are identical, within the measurement errors,  to that at 1$<$z$<$1.8.

\begin{figure}
\psfig{figure=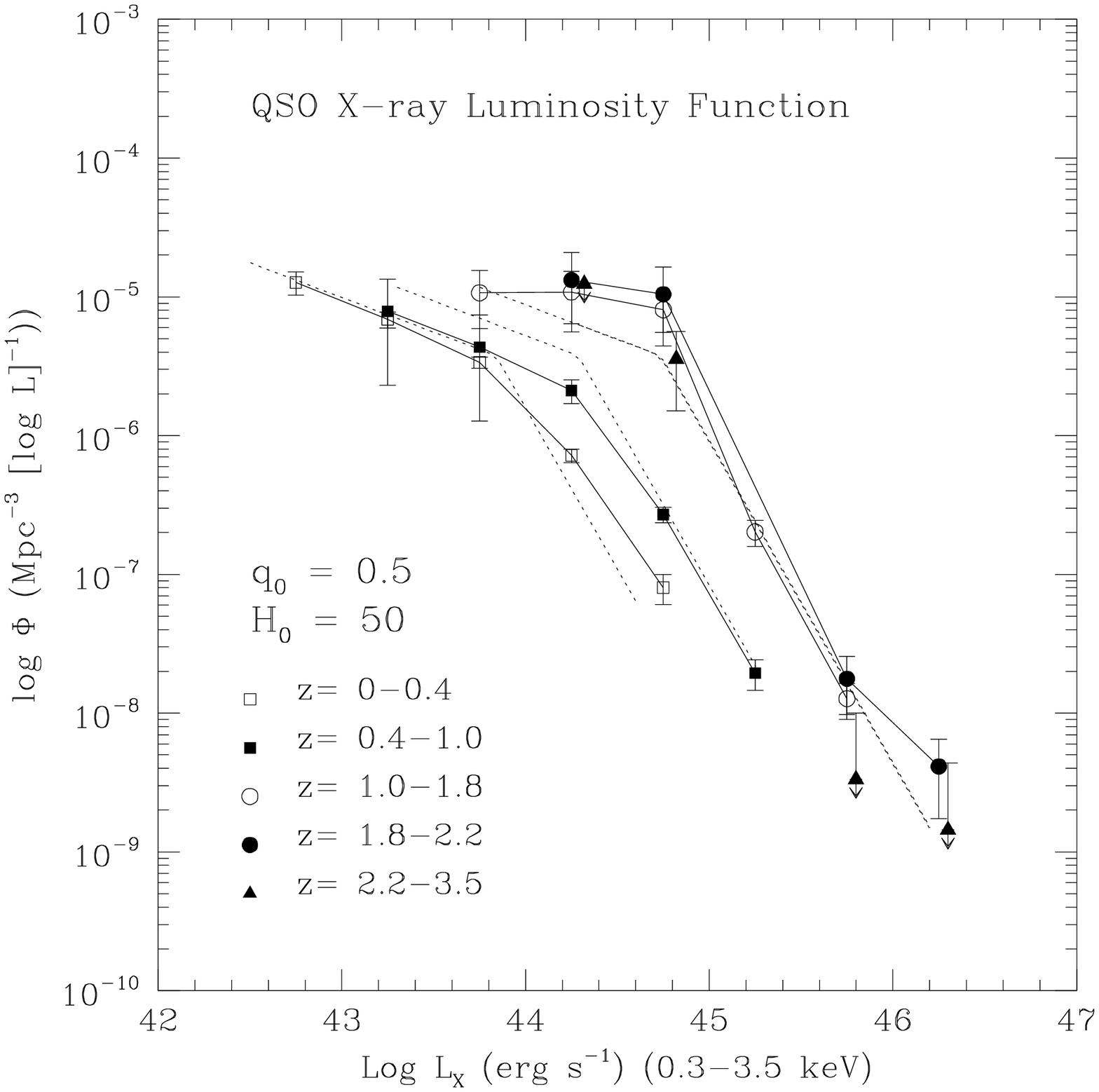,height=8.0truecm,width=8.0truecm,angle=0}
\psfig{figure=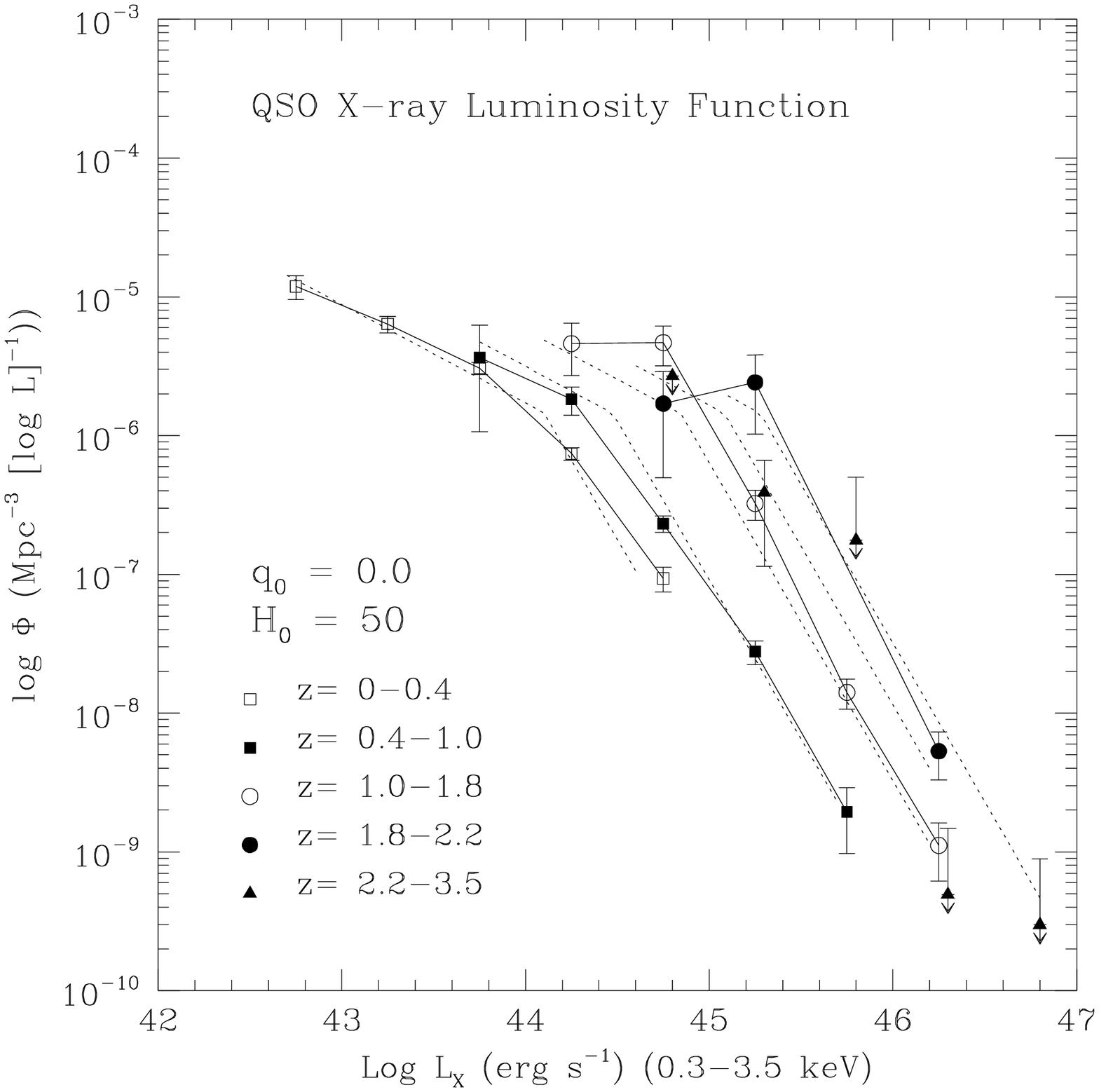,height=8.0truecm,width=8.0truecm,angle=0}
\caption{ 
Binned 1/V$_a$ estimates of the XLF in different redshift shells,
from the combined ROSAT and EMSS samples.
Upper limits are denoted by arrows. Some points have been moved slightly by 0.05 
in log(L$_X$) for clarity. See the text for details of the models.
(a) For q$_0$=0.5. Dotted lines are from model B. (b) For q$_0$=0. Dotted lines are from 
model D.
}
\end{figure}

The small number of QSOs detected at z$>$2.2 does not indicate a further change in the
evolutionary behaviour of the XLF beyond the halt in the luminosity 
evolution at z$\approx$1.8, despite the depth of the survey.
The filled triangles at L=10$^{44.25}$ erg s$^{-1}$ (Figure 5a) 
and L=10$^{44.75}$ erg s$^{-1}$ (Figure 5b) represent upper limits of 
$<$3 QSOs where none were detected. The upper limits are consistent with 
the XLF at lower redshifts 1.8$<$z$<$2.2.

The dotted lines in Figure 5a represent the best fit model of the XLF
(model B). At redshifts of 1$<$z$<$2.2 and luminosities of
10$^{44.5}<$L$<10^{45}$ erg s$^{-1}$ there is a small excess of QSOs
observed over the model prediction. This excess is not statistically
significant; 
2.5 are predicted and
5 were observed, a not unlikely 11\% probability.
This small excess is also visible in the 
n(z) data of Figure 2, where the excess of $\approx$3-4  QSOs over the 4 QSOs
expected at 1.5$<$z$<$1.75 has  $\approx$5\% chance of occurring in a
random distribution (again not unlikely given the 13 bins of Figure 2),
and at fluxes of 2-3x10$^{-14}$ erg cm$^{-2}$ s$^{-1}$ in the 
log(N)-log(S) relation of Figure 4b.  
The presence of large scale structure is not required to explain this
excess.


For q$_0$=0 (Figure 5b), there is much less evidence for a slowing of the 
luminosity evolution.
The 2.2$<$z$<$3.5 XLF seems to lie at approximately the same values as the 
1.8$<$z$<$2.2 XLF, but there are only 5 QSOs contributing to the 
2.2$<$z$<$3.5 XLF,
and the measurement errors are large. The apparent change in the slope of the
low luminosity part of the q$_0$=0 XLF with redshift (at luminosities less than
the break luminosity) is due to the small number
of QSOs at these luminosities at z$>$1.8. The observations at z$>$1.8 are
consistent within the measurement errors with the slope at low redshifts.

\subsubsection{Fitting an evolutionary model}

To derive more quantitative information, we used the maximum likelihood method
of Marshall \etal (1983)
to obtain the best-fit model of the shape of the XLF and its evolution.
We have investigated pure luminosity evolution models, guided by the 
appearance of the XLF in Figure 5, but not density
evolution models. The level of density evolution,
in which the XLF retains its shape but changes normalisation with redshift,
has been shown by Boyle \etal (1988) to be at least 60 times slower than the 
corresponding luminosity
evolution for high luminosity optically selected QSOs.
Following Maccacaro \etal (1991), Della-Ceca
\etal (1992) and Boyle \etal (1993, 1994) we define the differential XLF as a two power law
function with a break at luminosity L$^{*}_{z}$:

\[
\Phi(L)={dN\over{dLdV}}={{\Phi^*}\over{L^{*}_{{z}_{44}}}} \left( 
{{L_{44}}\over{L^{*}_{{z}_{44}}}} \right) ^{-\alpha_1} ~~~~~~~~~~~~ (L\leq{L^*_z})
\]
\[
\Phi(L)={dN\over{dLdV}}={{\Phi^*}\over{L^{*}_{{z}_{44}}}} \left( 
{{L_{44}}\over{L^{*}_{{z}_{44}}}} \right) ^{-\alpha_2} ~~~~~~~~~~~~ (L>L^*_z)
\]

The luminosities are in units of 10$^{44}$ erg s$^{-1}$ (0.3-3.5 keV).
The slopes at low and high luminosities are $\alpha_{1}$ and $\alpha_{2}$, and
the XLF normalisation is $\Phi^{*}$.
The normalisation is slightly different from that used by Boyle \etal (1993,1994):

\[
\Phi^*=\Phi^*_{B94} {L^{*}_{{z}_{44}}} ^{1-\alpha_1}
\]

We assumed a pure luminosity power law form
for the evolution of the XLF with a redshift cutoff z$_{cut}$:
\[
L^{*}_z(z) = L^{*}_0(0) (1+z)^{k}   ~~~~~~~~ (z<z_{cut})
\]
\[
L^{*}_z(z) = L^{*}_0(0) (1+z_{cut})^{k}   ~~~~ (z\geq{z_{cut}}).
\]
In this model the luminosity evolution continues up to redshift
z$_{cut}$, but beyond this redshift the luminosity function is invariant with
redshift. Evolutionary models both with and without this redshift cutoff
were investigated. Exponential luminosity evolution models were not investigated 
because they have been found to be a poor description of QSO X-ray evolution
by e.g. Boyle \etal (1993).

The number of free parameters to be constrained by the maximum likelihood fit
was four or five; $\alpha_1$, $\alpha_2$, L$^{*}_0$, k, and in some models, z$_{cut}$. 
The normalisation $\Phi^{*}$ was determined
by the total number of QSOs in the samples, and was not a free parameter. 

To estimate the goodness-of-fit
of the model to the data, the 2-dimensional Kolgorov-Smirnov (2D KS) test of
Fasano \& Franceschini (1987),
as implemented by Press \etal (1992), was used.
The test was performed over the complete range
of redshifts and luminosities available, 0$<$z$<$4 and 10$^{42}<$L$<10^{47}$ erg s$^{-1}$.
Having obtained a good fit, errors on the parameters were determined from 
the increase $\Delta$S in negative log(likelihood) 
away from the minimum value, allowing all free parameters to vary.
The value of $\Delta$S was chosen 
according to the number of free parameters
to give 68\% confidence limits (Lampton, Margon \& Bowyer 1976). 

\subsection{Results}
The results of the maximum likelihood fits and 2D KS tests are listed in Table 2.
For q$_{0}$=0.5, the power law luminosity evolution model with no redshift cutoff 
in the evolution is unacceptable
at the 1 per cent level (model A). Introducing the cutoff makes the model acceptable
at $>$20 per cent probability (model B). The best value for the redshift cutoff is
z$_{cut}$=1.41 (+0.4,-0.17, at 68\% confidence). The value of the evolution index
is k=2.97 (+0.19,-0.34).
These best fit values of z$_{cut}$ and k are in
excellent agreement with those found by Page \etal (1996) (z$_{cut}$=1.42, k=2.94),
and consistent with those found by Boyle \etal (1994)
based on ROSAT data alone (z$_{cut}$=1.6, k=3.25).

For q$_{0}$=0, an acceptable fit  is found for the power law luminosity evolution 
model at the 5 per cent confidence level with or without a redshift
cutoff (models D and C). The introduction of a redshift cutoff into the 
model gives a best fit value of z$_{cut}$=1.6 and increases the
best fit value of the evolution index from k=2.46 (+0.11,-0.14) to k=3.0 (+0.27,-0.4).

The ROSAT QSO log(N)-log(S) relation of Page \etal (1996) extended
to brighter fluxes than ours and overlapped with the EMSS relation. They found that
although the slopes of the two relations were consistent (and thus that the EMSS 
is probably reasonably complete), there was an offset in flux between the two relations.
This offset could be removed if a ROSAT 0.5-2 keV to {\it Einstein} 0.3-3.5 keV bandpass
conversion factor of 1.47 was adopted instead of the value of 1.8 given by a 
spectral index of 1.
We have investigated the effect of using this lower value in models I to L. 
There is no effect on which models are acceptable, although the probability
of both q$_0$ models is smaller, and the model parameters change
very little (see Table 2).

\begin{table*}
\begin{minipage}{175mm}
\caption {XLF and evolution parameters, EMSS+ROSAT data}
 \begin{tabular}{lclllllllll} \hline
Model & Sample(s)& q$_{0}$ & evolution & $\alpha_{1}$ & $\alpha_{2}$ & L$^{*}_{{0}_{44}}$ $^{\S}$ & k 
 & z$_{cut}$ & $\Phi_*$ $^{\dag}$ & P$_{KS}$ $^{\ddag}$ \\
 & & & & & & & & & & \\
            \hline
A &EMSS + ROSAT QSOs&0.5 & (1+z)$^k$     & 1.55  &       3.26 &  0.51 & 2.21 & - & 1.28 & 0.005 \\
B &"& 0.5 & (1+z)$^k$ ($<z_{cut}$) & 1.50$^{+0.20}_{-0.24}$  & 3.32$^{+0.17}_{-0.12}$ & 
 0.40$^{+0.17}_{-0.08}$ & 2.97$^{+0.19}_{-0.34}$ & 1.41$^{+0.40}_{-0.17}$ & 1.64 & 0.21 \\
C &"&0.0 & (1+z)$^k$    & 1.71$^{+0.11}_{-0.21}$ & 3.29$^{+0.12}_{-0.17}$ & 
 0.81$^{+0.18}_{-0.25}$ & 2.46$^{+0.11}_{-0.14}$ & - & 0.63 &  0.12 \\
D &"&0.0 & (1+z)$^k$ ($<z_{cut}$) & 1.58$^{+0.22}_{-0.25}$  & 3.31$^{+0.18}_{-0.13}$ &  
 0.50$^{+0.30}_{-0.12}$ & 3.03$^{+0.27}_{-0.40}$ & 1.61$^{+0.37}_{-0.27}$ & 1.07 & 0.045 \\
 & & & & & & & & & & \\
E &EMSS + ROSAT QSOs& 0.5 & (1+z)$^k$   & 1.72 & 3.32 & 
 0.6 & 2.31 & - & 1.00 & 0.010 \\
F & + ROSAT NELGs& 0.5 & (1+z)$^k$ ($<z_{cut}$) & 1.65$^{+0.24}_{-0.18}$ & 
 3.30$^{+0.34}_{-0.09}$ & 
 0.41$^{+0.42}_{-0.06}$ & 2.97$^{+0.24}_{-0.33}$ & 1.41$^{+0.39}_{-0.21}$ & 1.52 & 0.19 \\
G &"& 0.0 & (1+z)$^k$   & 1.84$^{+0.12}_{-0.12}$ & 3.30$^{+0.23}_{-0.11}$ & 
 0.88$^{+0.26}_{-0.22}$ & 
 2.46$^{+0.24}_{-0.14}$ & - & 0.53 & 0.17 \\
H &"& 0.0 & (1+z)$^k$ ($<z_{cut}$) & 1.78$^{+0.19}_{-0.18}$ & 3.29$^{+0.21}_{-0.10}$ & 
 0.64$^{+0.29}_{-0.21}$ & 
 2.79$^{+0.55}_{-0.14}$ & 1.70$^{+0.36}_{-0.35}$ & 0.76
 & 0.25 \\
 & & & & & & & & & & \\
I &EMSS + ROSAT QSOs& 0.5 & (1+z)$^k$   & 1.55 & 3.26 & 0.56 & 2.28 & - & 1.27 & 0.005 \\
J & S(0.3-3.5 keV)=& 0.5 & (1+z)$^k$ ($<z_{cut}$) & 1.50 & 3.30 & 0.40 & 2.97 & 1.37 & 1.63
 & 0.24 \\
K &1.47x S(0.5-2 keV)& 0.0 & (1+z)$^k$   & 1.64 & 3.08 & 0.70 & 2.20 & - & 0.83 & 0.053 \\
L &"& 0.0 & (1+z)$^k$ ($<z_{cut}$) & 1.59 & 3.25 & 0.56 & 2.79 & 1.66 & 0.97
 & 0.03 \\
 & & & & & & & & & & \\
M &EMSS only& 0.5 & (1+z)$^k$   & 1.71$^{+0.22}_{-0.25}$ & 3.33$^{+0.89}_{-0.24}$ & 
 0.78$^{+1.1}_{-0.23}$ & 
 2.18$^{+0.23}_{-0.23}$ & - & 0.77 & 0.07 \\
N &"& 0.5 & (1+z)$^k$ ($<z_{cut}$) & 1.62$^{+0.20}_{-0.29}$ & 3.31$^{+0.24}_{-0.27}$ & 
 0.50$^{+0.24}_{-0.16}$ & 
 2.80$^{+0.31}_{-0.28}$ & 1.41$^{+0.6}_{-0.21}$ & 1.17 & 0.29 \\
O &"& 0.0 & (1+z)$^k$   & 1.65$^{+0.20}_{-0.19}$ & 3.19$^{+0.32}_{-0.10}$ & 
 0.63$^{+0.31}_{-0.14}$ &  2.46$^{+0.24}_{-0.13}$ & - & 0.92 & 0.05 \\
P &"& 0.0 & (1+z)$^k$ ($<z_{cut}$) & 1.59$^{+0.24}_{-0.25}$ & 3.22$^{+0.29}_{-0.14}$ & 
 0.45$^{+0.5}_{-0.12}$ & 
 3.03$^{+0.23}_{-0.45}$ & 1.42$^{+0.8}_{-0.18}$ & 1.22 & 0.14 \\
 & & & & & & & & & & \\
             \hline
   \end{tabular}

$^{\S}$ Break luminosity at z=0 in units of 10$^{44}$ erg s$^{-1}$ (0.3-3.5 keV) \\
$^{\dag}$ XLF normalisation in units of 10$^{-6}$ Mpc$^{-3}$ (10$^{44}$ erg s$^{-1}$)$^{-1}$ \\
$^{\ddag}$ 2D KS probability\\
Errors, where quoted for models which are a good fit, are at 68\% confidence.

\end{minipage}
\end{table*}

\section{Discussion}
\subsection{Evolution models}
We have only considered pure luminosity evolution models. 
Pure density evolution would not match the binned XLF estimates of Figure
5.

For both q$_0$=0.5 and q$_0$=0, pure luminosity evolution is a good description of
the low redshift XLF evolution. The evolution can be characterized as a power law
in  (1+z), i.e. L$_{X}\propto{ (1+z)^{3.0 (+0.19,-0.34)} }$. 
For q$_0$=0.5, this evolution halts at a redshift
of 1.4 (+0.4,-0.17), and we detect no further change in the XLF up to z$\sim$3.
The log(N)-log(S) prediction of this model (model B) is shown as a solid
line in Figure 4. Figure 4a shows that the model predicts 
a lower number of ROSAT QSOs than were observed, but the prediction of the total
number of QSOs in the ROSAT survey
is only 1.5$\sigma$ below the total number observed. The 
integral log(N)-log(S) plot can be misleading since the data points are 
not independent; the model is consistent 
with the differential log(N)-log(S) relation shown in Figure 4b. 
In three surveys (this work, Page \etal 1996 and the ROSAT data of 
Boyle \etal 1994) it has now been found that,
for q$_0$=0.5, the luminosity evolution of the X-ray QSO XLF halts at a redshift of
z=1.4-1.6 and the evolution parameter k lies in the range k=2.9-3.25
(where the comparison is with the values obtained using a ROSAT to EMSS bandpass
conversion factor of 1.8).

For q$_0$=0, although a halt in the luminosity evolution is not required by the 
data, when it is included (model D), the value of z$_{cut}$ is 1.6,
again similar to 
the values found by Page \etal (1996) (1.82) and Boyle \etal (1994) (1.79 for
ROSAT data alone). The evolution parameter k is 3.0, 2.9 and 3.3 in
this work, Page \etal and Boyle \etal respectively. The log(N)-log(S) prediction of
model D, shown as a dashed line in Figures 4a and 4b, matches the
observed log(N)-log(S) relations and predicts more faint QSOs than
model B (for which q$_0$ is 0.5). The n(z) prediction of model D,
using the normalisation given in Table 2, is shown in Figure 2. The prediction
matches the observed QSO n(z) distribution. 

\subsection{Errors of derived parameters}
Before discussing in detail the different results of ROSAT, EMSS
and optical surveys, we examine the methods used to determine the
size of the errors  of the parameters describing the shape
and evolution of the XLF, since these are crucial in interpreting
different results.
We estimated the errors by
stepping each parameter away from the best-fit value, reminimizing S (where
S=-2ln($L$) and $L$ is the likelihood function), and finding the parameter value
which gave an increase in S of $\Delta$S above the minimum. The appropriate
value of $\Delta$S to use is a function of the confidence level required and the number
of free parameters (Lampton, Margon and Bowyer 1976), since the error values
obtained are a projection of the multi-dimensional confidence region on to each
parameter axis in turn. For 68\% (90\%) confidence and 4 free parameters, 
we used $\Delta$S=4.7 (7.78);
for 5 free parameters, we used $\Delta$S=5.9 (9.2). 
Since the errors on different parameters are correlated, a parameter value at the 
edge of a confidence region as defined here will not in general be consistent with
the full confidence region of other parameters, but only a smaller region.
However, our definition of the confidence region does give an estimate of
the total range of values each parameter could take, taking into account
the errors on all the other parameters.
The errors are in general non-linear
and non-symmetric, so the 90\% confidence errors for many parameters are only
slightly larger than the 68\% confidence errors. As an consistency check, we
noted the values of the 2D KS probability P$_{KS}$ as
z$_{cut}$ was stepped  away from the best-fit value of 1.41 in model B, and
S reminimized at each step. P$_{KS}$ fell below 0.1 (i.e. 90\% probability) 
at 1.41 (+0.5,-0.4), in approximate agreement with the $\Delta$S=9.2 (90\%) values
of +0.5,-0.21 and $\Delta$S=5.9 (68\%) values of +0.4,-0.17.


Some previous QSO LF investigations (Page \etal 1996; Boyle \etal (1993,1994); Boyle, Shanks 
\& Peterson 1988) have used $\Delta$S=1 when models have had 4 or more free parameters.
As noted by Boyle, Shanks \& Peterson (1988), $\Delta$S=1 is only strictly valid 
for the case of one parameter taken in isolation (e.g. the confidence region 
for the evolution parameter k assuming zero error on the XLF shape and
redshift cutoff).
For our combined ROSAT+EMSS surveys and models, a value of $\Delta$S=1
gives errors smaller by factors varying from 2 to 5 compared with 
the higher $\Delta$S values used here. Boyle \etal (1994)
note that a better estimate of their parameter errors may be obtained from
the variations in parameter values from model to model rather than the $\Delta$S=1
values. A comparison can also be made with the error estimates of Maccacaro \etal (1991) 
and Della-Ceca \etal (1992), who analysed the EMSS using a different method from
that used here. Their 68\% confidence error estimates of k=2.56$\pm$0.17, derived 
from the error on the V/V$_{max}$ statistic, are in reasonable agreement with 
our analysis of the same data (model O) which gives k=2.46 (+0.24,-0.13).

\subsection{ROSAT and EMSS X-ray QSO evolution}
Previous authors have emphasized the different values obtained for the pure luminosity
evolution 
parameter k from the EMSS and ROSAT surveys. The EMSS value of k=2.56$\pm$0.17
 (for q$_0$=0; Della-Ceca \etal 1992) 
appears significantly lower than the ROSAT value of 3.34$\pm$0.1 (Boyle \etal 1994) and
the combined ROSAT+EMSS values of 2.9-3.0 (this paper, Page \etal 1996 and Boyle 
\etal 1994).
Franceschini \etal (1994) offered a possible explanation by analysing a subset of 
the EMSS data, restricted
to fluxes $>$2.6x10$^{-13}$ erg cm$^{-2}$ s$^{-1}$, and finding a best-fit value of
k=3.27 (assuming an X-ray spectral index of one). They suggested that the EMSS might be incomplete 
at faint fluxes.

However, it is important to compare results from the same assumed evolutionary model.
Our result (for the combined ROSAT+EMSS data and q$_0$=0) that the value of k 
increased significantly when a redshift cutoff was introduced into the model
prompted us to re-analyse the EMSS data including a redshift cutoff. This 
model was not considered by Maccacaro \etal (1991) or Della-Ceca \etal (1992).
The results are given in Table 2, models M-P. For q$_0$=0 and no redshift 
cutoff (model O), k=2.46 (+0.24,-0.13), consistent with the previous EMSS results.
When a redshift cutoff is introduced, the evolution index increases to 
k=3.03 (+0.23,-0.45), and the redshift cutoff value
is z$_{cut}$=1.4 (+0.8,-0.18) (model P). The large errors are an indication
that model P is over-complicated and the EMSS data do not require a 
redshift cutoff; nevertheless, if a redshift cutoff is included in the model,
the EMSS data are best fit with a high value
of k, consistent with the results from combined EMSS+ROSAT surveys, and the
ROSAT data alone of Boyle \etal (1994). In addition, this model, derived
from the EMSS data alone, predicts a redshift distribution for the ROSAT survey
which is consistent with the ROSAT data (see Figure 2).
A similar increase in
the best-fit EMSS value of k when a redshift cutoff is included, from k=2.18 to k=2.80, 
is found for q$_0$=0.5 (models M and N). 

The converse is also in general true.
If a redshift cutoff is {\it not} included, the combined
ROSAT+EMSS surveys of this paper, Page \etal (1996) and Boyle \etal (1994) give
consistent values of k=2.46 (+0.11,-0.14), 2.66$\pm$0.08 and 2.63$\pm$0.1
respectively (q$_0$=0). 
Ciliegi  \etal (1995) also found relatively low values of the evolution parameter
from combined ROSAT and EMSS samples, with no redshift cutoff in their model.
The ROSAT data alone of Boyle \etal (1994) give k=2.66$\pm$0.1.
Boyle \etal (1993), with a smaller number of ROSAT QSOs, found a value of 
k=2.75$\pm$0.1 for their combined ROSAT+EMSS survey and k=2.9 for the ROSAT data 
alone. The value of k=2.75$\pm$0.1 is consistent with the EMSS value of 2.56$\pm$0.17
at the 68\% confidence level, and the addition of more ROSAT QSOs (Boyle \etal 1994)
has resulted in a new value of 2.63$\pm$0.1.

The reason for the increase in the EMSS evolution parameter when a redshift cutoff
is introduced is that many of the EMSS QSOs are at redshifts higher than the 
redshift z$\approx$1.6 where the evolution halts or slows (see Figure 3). 
The evolution parameter is insensitive to the numerous low redshift 
z$\sim$0.2 EMSS QSOs, so if a single value of k is assumed to apply at all
redshifts, with no redshift cutoff, the high redshift QSOs weight the best fit of a straight
line in the log(L$^*_z$)-log(1+z) plane to a flatter slope, or a lower value of k.

Thus, although the EMSS contains too few high redshift QSOs to constrain
the redshift cutoff, the EMSS data are consistent with an evolution index of
k=3.0,
higher than the value of k=2.56 found by Maccacaro \etal (1991) and Della Ceca \etal
(1992), and consistent with that found from recent ROSAT surveys. 
In general, four X-ray surveys (the EMSS and those of Boyle \etal (1994), Page \etal (1996) 
and this work) give consistent results for the evolution index: k=2.8-3, 
with a redshift cutoff of z$_{cut}$=1.4-1.8. There are, however, two caveats
to this statement. First, none of the surveys is completely independent, since
all incorporate the EMSS. Boyle \etal (1994) found a slightly higher value of k=3.2-3.3
from ROSAT data alone (with a fixed XLF shape). Secondly, Boyle \etal (1994) 
also required more complicated models (e.g. a `polynomial' 
pure luminosity evolution model) 
in order to achieve statistically acceptable fits to the combined 
ROSAT and EMSS data, although each dataset could be fit separately by the simpler
luminosity evolution models used here. A possible explanation
is investigated in section 4.5.

The high value of the EMSS evolution parameter found by Franceschini \etal (1994)
was obtained with a redshift cutoff fixed at z$_{cut}$=2.5 and a bright flux limit. 
The bright flux limit will have reduced the number of high redshift QSOs
somewhat; the combination of this factor and the redshift cutoff (although it
has a high value) may have partly produced the increase in the evolution parameter,
rather than incompleteness in the EMSS, as suggested by Franceschini \etal (1994).

\subsection{X-ray and optical QSO evolution}
Comparing X-ray and optically selected QSO surveys, the X-ray evolution index 
(k=2.97 +0.19,-0.34)
and redshift cutoff (z$_{cut}$=1.4 +0.4,-0.17)
found here for q$_0$=0.5 are both lower than that found by Boyle \etal (1991)
(k=3.45, z$_{cut}$=1.9) from a combination of optical surveys. Although the values of 
z$_{cut}$ are consistent at 68\% significance, this is a general
finding for all X-ray QSO surveys. These differences cannot arise from different sampling of
redshift space assuming the simple evolution model used here; if the X-ray sampling
and measurement errors produced a best-fit redshift cutoff at a lower redshift
than the true value, then a higher value of the X-ray evolution index would be produced,
not a lower value. However, recent optical surveys indicate that simple pure
luminosity evolution may no longer be an adequate description of optical QSO evolution.
Hewett \etal (1993), in an analysis of an independent optical survey, find that
optical luminosity evolution slows down at a redshift of z$\sim$1.5, and that it continues 
at a slower rate  corresponding to k$\sim$1.5 up to a redshift of z$\sim$3. 
In addition, Hewett \etal (1993) and Miller \etal (1993) observe a change in
shape of the optical luminosity function with redshift, such that the high luminosity
slope is steeper at higher  redshift. Neither of these effects are currently  observed in
X-ray samples. 


\subsection{Absorption and emission features in QSO X-ray spectra}

The range of rest frame energies in the ROSAT and EMSS surveys is
0.3-3.5 keV for the EMSS at z=0, 2-8 keV for ROSAT at z=3, and 0.9-10.5 keV 
for the EMSS at z=2.  The 
assumption of a mean QSO spectrum of a smooth power law of spectral
index one is unlikely to be correct. Here we estimate how good
that assumption is. These estimates do not include the details
of the different instrumental responses, but rather provide an
approximate estimate of the size of the effects.

Although the value of the mean spectral index $\approx$1 obtained from
low resolution, low signal-to-noise X-ray spectra (typical of those from
ROSAT surveys and the EMSS) is confirmed by detailed studies of brighter
AGN (e.g. Nandra \& Pounds 1994), the observed spectra contain emission
lines and photoelectric absorption features. The potential effects are on
the count rate to flux conversion factor, the ROSAT-{\it Einstein} bandpass
conversion factor, and the K-correction. In practice, the effect on the
count rate to flux conversion is negligible, as shown by the wide range 
of power law indices that have $<$10\% effect (see section 3.1).

An absorption edge and fluorescent emission line (at a rest energy of 
6.4 keV) due to
highly ionized iron, together with scattered photons at higher
energies (the  `reflection bump') have been found to be common in
Seyfert I AGN (Nandra \& Pounds 1994). The iron features are redshifted
into the ROSAT band at z$>$2.2 and into the EMSS band at z$>$0.9.
However, the typical iron line equivalent width of $\approx$200 eV 
would produce a change in the flux in the EMSS or ROSAT bands of
$\leq$2\%, small enough to ignore. The reflection bump, at rest energies 
$>$8 keV, will have a larger
effect on ROSAT QSOs at z$>$3 and EMSS QSOs at z$>$1.3. For the highest
redshift  EMSS QSOs (at z$\sim$2.5) the effect of the reflection bump
would be to change the EMSS flux by $<$10\%. However, EMSS QSOs at these
redshifts are
all of high luminosity ($>$10$^{45}$ erg s$^{-1}$), and there have
been fewer detections of a reflection bump in high luminosity sources
compared to lower luminosity AGN (e.g. Williams \etal 1992).

A more important feature has been detected in some ASCA AGN spectra.
The ASCA spectral resolution has revealed what may be the strongest
feature in the 0.4-6 keV band; OVII and OVIII absorption edges at
0.72 keV and 0.87 keV, known as `warm' absorbers (e.g.  Otani \etal 1996a,
Fabian \etal 1994 and 
references therein). Otani \etal (1996b) found that six out of eight
type 1 AGN, with luminosities of 10$^{42}$ to 10$^{45}$ erg s$^{-1}$,
contain similar absorption features. The total  absorption was $\approx$20\%
of the 0.5-2 keV flux in all six objects, and was all within the 0.5-2 keV
band. The fraction of AGN which exhibit such absorption features
is not well known, although Nandra \& Pounds (1994) found $\sim$50\% 
of their sample of 28 AGN to do so. We will estimate the worst-case 
systematic effect of 
all AGN containing a warm absorber. 
The effect on the ROSAT 0.5-2 keV
to {\it Einstein} 0.3-3.5 keV
passband conversion at zero redshift is to increase the factor 
from 1.8 to $\approx$2.0, since a smaller fraction of the {\it Einstein} flux is
absorbed. The size of this effect is similar to that required by
Page \etal (1996) to match ROSAT and EMSS log(N)-log(S) relations,
but in the opposite sense; Page \etal required a conversion factor
of 1.47. At z$>$1.6, the absorption features are redshifted 
out of the ROSAT band, but may effect the EMSS fluxes at energies below
the carbon edge in the detector windows, given the modest energy resolution. 
Whilst the exact value of the conversion factor depends on the instrumental
responses and resolutions, a value smaller than 1.8, perhaps $\approx$1.6,
may be required for some of our ROSAT QSOs.
We have already shown, however, that
decreasing the conversion factor from 1.8 to 1.47 (models I-J)
has little effect on our results. The redshift-dependent effect of the warm absorber
spectral feature may however help explain the ROSAT/EMSS log(N)-log(S) 
discrepancy of Page \etal (1996), and partly explain the poor fits  
found by Boyle \etal (1994) in their combined EMSS+ROSAT sample, with
a much larger number of ROSAT QSOs than used here.

The effect of the warm absorbers on the K-correction in the 
ROSAT band would be to 
decrease the K-correction with increasing redshift from 1 at low redshifts
to a minimum of $\approx$0.8 at high redshifts (z$\ga$1).
In the EMSS band, the minimum K-correction would be 
$\approx$0.9. These changes
to the K-corrections are small compared to the luminosity evolution observed 
(e.g. for L$\propto$(1+z)$^3$, a change in luminosity by a factor
of nine from z=0.4 to z=2). Thus we would expect 
the detailed values of the XLF parameters to change slightly,
but not the general conclusions, including which evolutionary models produce
acceptable fits to the data.

\subsection{Narrow Emission Line Galaxies}
Sixteen of the ROSAT X-ray sources have been identified with narrow 
emission line galaxies (NELGs) by \mch\etal (1996). From optical emission
line diagnostics, some may contain
absorbed AGN. Since the EMSS also contains examples of this class of source,
and these are included in the EMSS AGN sample we have used here, we have
investigated the effect of including the ROSAT NELGs in the analysis.
All the NELGs, in both ROSAT and EMSS samples, lie at redshifts
z$<$0.6, so their inclusion is unlikely to affect the high redshift XLF behaviour.
The luminosities of the 
ROSAT NELGs lie in the range 3x10$^{41}$ - 10$^{43}$ erg s$^{-1}$, but 
we have only included NELGs of luminosity $>10^{42}$ erg s$^{-1}$ in the analysis, 
in order to maintain consistency with the previous analysis which used
a lower limit of 10$^{42}$ erg s$^{-1}$ in the maximum likelihood calculation,
and because galaxies
below this luminosity are more likely to be normal or star-forming galaxies, not containing
AGN.
Above this luminosity, there were 12(13) NELGs for q$_0$=0.5(0). 

The results are given in Table 2, models E to H. In general, the only effect
is to increase the  
low luminosity slope of the XLF by $\approx$0.15 (not surprising given the 
low luminosities of the NELGs). All other parameters are 
virtually unchanged. A redshift cutoff in the 
luminosity evolution at z$_{cut}\approx$1.4 is still required for q$_0$=0.5.

The nature of the NELGs, and the origin of their X-ray emission, is not well
understood at present. The redshift distributions of the NELGs and the QSOs are
clearly very different; the NELG+QSO n(z) is plotted
as a dashed line in Figure 2. We argue here that the NELG population cannot consist
completely of low luminosity AGN because of the double-peaked shape of this 
n(z) distribution. Although the evolutionary models with a 
relatively steep low luminosity XLF slope produce 
acceptable fits to the ROSAT and EMSS data combined, they do not produce a double
peaked redshift distribution, as shown by the n(z) prediction of model H in 
Figure 2.
A one-dimensional KS test on the unbinned data finds the  n(z) distribution of 
the combined ROSAT QSOs and NELGs only marginally consistent with the prediction of model H
at the 4\% level, whereas the n(z) distribution of the ROSAT QSOs alone is consistent 
with the prediction of model D at the $>$20\% level.
In order to produce a double peaked QSO redshift distribution, a re-steepening of the 
low luminosity XLF slope would be required at the lowest luminosities,
beyond the `flat' portion of the XLF (L$\la$10$^{43}$ erg s$^{-1}$ and
z$\sim$0.3). No such re-steepening was observed by Maccacaro \etal (1991) in
their 0$<$z$<$0.18 EMSS XLF which extended to luminosities of 10$^{42}$ erg s$^{-1}$, and
included the EMSS NELGs. The NELGs may have highly absorbed X-ray spectral
components which are undetected in soft X-ray surveys. 
However, no evidence for a large column density
($>$10$^{21}$ cm$^{-2}$) 
in the ROSAT NELG spectra was found by Romero Colmenero \etal (1996) and
the relatively faint optical fluxes of the NELGs do not imply very large X-ray fluxes
(\mch\etal 1996).
A simpler explanation is that some fraction
of the NELGs, or some fraction of their X-ray emission, is from a different, 
non AGN, origin. 

\section{The QSO contribution to the soft X-ray background}
In order to estimate the total QSO contribution to the 1-2 keV X-ray background (XRB), 
including the contribution 
from QSOs of flux fainter than the limit of the ROSAT survey, we integrated the evolving QSO
XLF, including luminosities and redshifts which have not been directly sampled by any
survey. 
The intensity of the 1-2 keV extragalactic XRB was taken to be 1.25x10$^{-8}$ erg
cm$^{-2}$ s$^{-1}$ sr$^{-1}$, as measured by Hasinger \etal
(1993) and confirmed by Gendreau \etal (1995). The integration was performed over
the redshift range 0$<$z$<$4 and over the fixed luminosity range 10$^{42}<L<10^{48}$
erg s$^{-1}$. Since the XLF was measured in the 0.3-3.5 keV band, the 
values of the integrated QSO surface brightness, $I_Q$, were multiplied by 0.28
to convert to the 1-2 keV band, assuming an X-ray spectral index of 1. 
The values of $I_Q$ and the corresponding XRB fractions are given in table 3 
for various models which give good evolutionary fits.
The 1-2 keV XRB fractions range from 31\% to 51\%.

The XRB fraction depends most strongly on the assumed cosmology; the values for
q$_0$=0 are $\approx$50\% and those for q$_0$=0.5 are $\approx$35\%. Smaller
differences are caused by assuming that all of the X-ray luminosity of the ROSAT
NELGs arise in AGN, and including them in the analysis (models F and H), or by the
existence or not of a halt in the XLF luminosity evolution (models C and D).

The lower luminosity limit of 10$^{42}$ erg s$^{-1}$ was chosen because this
is the lower limit of the observed XLF at redshift zero. The surface density
of QSOs contributing to the XRB using this lower limit varies from 2000 deg$^{-2}$ 
to 30000  deg$^{-2}$ depending on the model. 
Decreasing the lower luminosity limit to 10$^{40}$ erg s$^{-1}$
increases the XRB contributions of most of the models by 2-4\%, up to a
maximum of 9\% for model H, and increases the assumed QSO surface density to 
2x10$^{4}$ deg$^{-2}$ 
to $\sim$10$^{6}$  deg$^{-2}$, of the same order as the total surface density of all the faintest galaxies
currently observed. At these low luminosities, the AGN luminosity would in any case
be similar to the host galaxy X-ray luminosity.
 The sensitivity of model H, which includes the ROSAT NELGs, 
to the lower luminosity limit is because model H has the steepest
value of the low luminosity XLF slope ($\alpha_1$=1.78) of any of the models considered
here. Apart from the assumed cosmology, the uncertainty in the low luminosity XLF 
slope ($\alpha_1$) contributes the largest 
uncertainty of any of the XLF or evolutionary paramters. The uncertainty in the XRB
contribution corresponding to the 68\% confidence range in $\alpha_1$ for model B,
leaving the other parameters unchanged, 
is 31\% (+9\%,-7\%) for a lower luminosity limit of 10$^{42}$ erg s$^{-1}$, and 
31\% (+15\%,-7\%) 
for a lower limit of 10$^{40}$ erg s$^{-1}$.

A recent measurement of the absolute 1-2 keV extragalactic XRB intensity 
incorporating ASCA and ROSAT data gives a value of 1.46x10$^{-8}$ erg cm$^{-2}$ s$^{-1}$
sr$^{-1}$ (Chen, Fabian \& Gendreau 1996). Using this higher value would decrease
the measured QSO contributions from 31\%-51\% to 27\%-44\%.

\begin{table}
\centering
\caption {QSO contribution to the 1-2 keV XRB.}

\begin{tabular}{llll} \hline
 Model & q$_0$& $I_Q$, 10$^{-8}$ erg cm$^{-2}$ s$^{-1}$ sr$^{-1}$ & XRB fraction \\
       &  & (0.3-3.5 keV) & (per cent) \\
\hline
 & & & \\
B & 0.5  & 1.37 & 31\% \\
C & 0.0  & 2.23 & 50\% \\
D & 0.0  & 2.05 & 46\% \\
F & 0.5  & 1.58 & 35\% \\
H & 0.0  & 2.28 & 51\% \\
\end{tabular}
\end{table}

\section{Conclusions}
We have performed a very deep ROSAT X-ray survey and measured the highest yet
QSO surface density of 230$\pm$40 deg$^{-2}$. A break in the QSO log(N)-log(S) 
relation has been found at a flux of $\approx$2x10$^{-14}$ erg cm$^{-2}$ s$^{-1}$
(0.3-3.5 keV). We have shown that 
measurement of the QSO XLF at luminosities fainter than the break in the XLF
at z$>$2.2 is possible with very deep ROSAT surveys. 

From a combination of the {\it Einstein} EMSS and our deep ROSAT survey, we find that pure 
luminosity evolution of the form L$_{X}\propto$(1+z)$^{3.0(+0.2,-0.3)}$ is a
good description of
the QSO XLF evolution at low redshifts. This evolution is consistent with 
that found in previous X-ray surveys, including from the EMSS alone, when
the same evolutionary model is used in the comparison. 

At higher redshifts, we find that a halt, or strong slowing, of the luminosity 
evolution is required at a redshift of z=1.4 (+0.4,-0.17) (for q$_0$=0.5). 
For q$_0$=0, the combined surveys are consistent with a halt to the evolution 
at z=1.6 (+0.37,-0.27), but the data do not require the halt. These conclusions
are still valid if all the narrow emission line galaxies found in the ROSAT survey
are assumed to be absorbed AGN. In fact,
the very different redshift distributions of the QSOs and the narrow emission
line galaxies suggest that some fraction of the 
X-ray flux from the narrow emission line galaxies arises from a non-AGN source. 

Unexplained differences in the evolution parameters derived from optically selected
surveys (e.g. k=3.45, z$_{cut}$=1.9, Boyle \etal 1991) and X-ray selected surveys
exist. The change in slope with redshift of the high luminosity part of the optical LF 
observed in recent optical surveys is also not observed in X-ray surveys.
Accurate treatment of the errors on the evolutionary parameters and future
X-ray samples (both shallow and deep) will help resolve the differences.

We find that QSOs contribute between 31\% and 51\% of the 1-2 keV extragalactic
X-ray background. Finally,
we note that the errors on the 
parameters describing the luminosity function and its evolution
may have been underestimated in some previous surveys.

\section{Acknowledgements}
We thank Mike Irwin for obtaining APM measurements, and acknowledge discussion
of AGN X-ray spectra with Paul Nandra.
We also thank the following observatories and their staff for support of
this project: the Canada-France-Hawaii Telescope, the UK William
Herschel Telescope, the Nordic Optical Telescope, the University of
Hawaii 88 inch Telescope the Mitchigan-Dartmouth-MIT Telescope and the
Very Large Array Radio Telescope.  This work was supported by grants
to a number of authors from the UK Science and Engineering Research
Council and Particle Physics and Astronomy Research Council. KOM
acknowledges support from the Royal Society. Summer student Scott Moore 
performed some of the CCD data reduction.\\

{\bf REFERENCES}\\

\noindent
Avni, Y., \& Bachall, J.N., 1980, ApJ, 325, 694.\\
Barcons, X., Branduardi-Raymont, G., Warwick, R.S., Fabian, A.C., 
   Mason, K.O., M$\rm^{c}$Hardy, I.M., Rowan-Robinson, M., 1994,
   MNRAS, 268, 833.\\
Boyle, B.J.,  Shanks, T., Perterson, B., 1988, MNRAS, 238, 957.\\
Boyle, B.J., Jones, L.R., Shanks, T., Marano, B., Zitelli, V., Zamorani, G.,
 1991, in "The Space Distribution of Quasars",
 ASP Conference Series Vol 21, ed Crampton, D.\\
Boyle, B.J.,  Griffiths, R.E., Shanks, T., Stewart, G.C., Georgantopoulos, I.,
 1993, MNRAS, 260, 49.  \\
Boyle, B.J., Shanks, T., Georgantopoulos, I., Stewart, G.C.
  Griffiths, R.E., 1994, MNRAS 271, 639.\\
Boyle, B.J., McMahon, R.G., Wilkes, B.J., Elvis, M., 1995, MNRAS, 272, 462.\\
Branduardi-Raymont, G. \etal 1994, MNRAS, 270, 947.\\
Briel, U.G. \etal 1995, "The ROSAT Users Handbook", MPE.\\
Cash, W., 1979, ApJ 228, 939. \\
Chen, L.-W., Fabian, A.C., Gendreau, K.C., MNRAS, in press.\\
Ciliegi, P., Elvis, M., Wilkes, B.J., Boyle, B.J., McMahon, R.G.,
Maccacaro, T., 1995, MNRAS, 277, 1463.\\
Della Ceca, R., Maccacaro, T., Gioia, I.M., Wolter, A.,
 Stocke, T.J., 1992, ApJ, 389, 491.\\
Fabian, A.C., \etal 1994, PASJ, 46, L59.\\
Fasano, G., Franceschini, A. 1987, MNRAS, 225, 155.
Franceschini, A., La Franca, F., Cristiani, S., Martin-Mirones, J.M., 
 1994, MNRAS, 269, 683.\\
Fruscione, A., Griffiths, R.E., 1991, ApJ 380, L13.\\
Gendreau, K.C., \etal 1995. PASJ, 47, 5.\\ 
Halpern, J.P., Helfand, D.J., Moran, E.C., 1995, ApJ, 453, 611.\\
Hasinger, G., Burg,R., Giacconi, R., Hartner, G., Schmidt, M., Trumper,
    J., Zamorani, G., 1993, A\&A 275, 1.\\
Hasinger, G., Boese, G., Predehl, P., Turner, T.J., Yusaf, R., George, I. \&
 Rohrbach, G., 1993b, MPE/OGIP Calibration Memo Cal/Ros/93-015.\\
Hewett, P.C., Foltz, C.B., Chaffee, F.H. 1993, ApJ, 406, L43.\\
Jones, L.R., \etal 1994, Proc. of
   35th Herstmonceux Conference, eds Maddox, S. and Aragon-Salamanca, A.,
   World Scientific Publishing, p.339.   \\
Lampton, M. Margon, B., Bowyer, S. 1976, ApJ 208, 177.\\
M$\rm^{c}$Hardy, I.M., \etal 1996, MNRAS submitted.\\
Maccacaro, T., Della Ceca, R., Gioia, I.M., Morris, S.L., Stocke, J.T.,
    Wolter, A., 1991, ApJ, 374, 117.\\
Marshall, H.L., Avni, Y., Tananbaum, H., Zamorani, G., 1983, ApJ,
 269, 35.\\
Metcalfe, N., Shanks, T., Fong, R., Jones, L.R., 1991, MNRAS, 249, 
   498.\\
Miller, L., Goldschmidt, P., La Franca, F., Cristiani, S., 1993, in Observational
 Cosmology, ASP Conf. Ser. vol 51, eds, G. Chincarini, \etal. \\
Nandra, P., Pounds, K.A., 1994, MNRAS, 268, 405.\\
Otani, C. \etal 1996a, PASJ, in press.\\
Otani, C. \etal 1996b, Proc. "Roentgenstrahlung from the Universe", in press.\\
Page, M.J., Carrera, F.J., Hasinger, G., Mason, K.O., McMahon, R., Mittaz, J.P.D.,
 Barcons, X., Carballo, R., Gonzalez-Serrano, I., Perez-Fournon, I., 1996, MNRAS,
 in press. \\
Press, W.H., Teukolsky, S.A., Vetterling, W.T. \& Flannery, B.P. 1992,
 Numerical Recipes, 2nd edition, Cambridge University Press.\\
Romero Colmenero, E., Branduardi-Raymont, G., Carrera, F.J., Jones, L.R.,
 Mason, K.O., McHardy, I.M., Mittaz, J.P.D., 1996, MNRAS, in press.\\
Shanks, T., Georgantopoulos, I., Stewart, G.C., Pounds, K.A., Boyle, B.J. 
   and Griffiths, R.E. 1991. Nature, 353, 315. \\ 
Stocke, J.T., Morris, S.L., Gioia, I.M., Maccacaro, T., Schild, R.,
    Wolter, A., Fleming, T.A., Henry, J.P., 1991, ApJS, 76, 813.\\
Wilkes, B.J., Elvis, M., 1987, ApJ, 323, 243.\\
Williams, O.R. \etal 1992, ApJ, 389, 157.\\



\end{document}